\newlength{\extralineskip}
\documentstyle[12pt]{article}
\begin{document}
\begin{titlepage}
\begin{flushright}
          \begin{minipage}[t]{12em}
          \large UAB--FT--366\\
                 May 1995
          \end{minipage}
\end{flushright}

\vspace{\fill}

\vspace{\fill}

\begin{center}
\baselineskip=2.5em

{\LARGE  THE ORTHOPOSITRONIUM DECAY PUZZLE \\
AND PRIMORDIAL NUCLEOSYNTHESIS}
\end{center}

\vspace{\fill}

\begin{center}
{\sc R. Escribano, E. Mass\'o and R. Toldr\`a}\\

     Grup de F\'\i sica Te\`orica and Institut de F\'\i sica
     d'Altes Energies\\
     Universitat Aut\`onoma de Barcelona\\
     08193 Bellaterra, Barcelona, Spain
\end{center}

\vspace{\fill}

\begin{center}

\large ABSTRACT
\end{center}
\begin{center}
\begin{minipage}[t]{36em}
The discrepancy between the experimental decay rate of
orthopositronium (o-Ps) and the QED theoretical
prediction can be solved by invoking decays of o-Ps
into exotic particles with branching ratios
$\sim 10^{-3}$. We show that considerations
based on primordial nucleosynthesis and effective
Lagrangians place a very stringent upper bound:
$B\equiv \Gamma(\mbox{o-Ps}\rightarrow ``exotic"+...)/
\Gamma(\mbox{o-Ps}) \leq 2 \times 10^{-15}$, ruling out the
exotic decay solution to the puzzle.
\end{minipage}
\end{center}

\vspace{\fill}

\end{titlepage}

\clearpage

\addtolength{\baselineskip}{\extralineskip}
\newcommand{\oPs}{\rm o-Ps}
\newcommand{\keV}{\rm keV}
\newcommand{\second}{\rm s}
\newcommand{\MeV}{\rm MeV}
\newcommand{\GeV}{\rm GeV}

{\bf 1.Introduction}

In the last few years, experiments \cite{Westbrook87&89,Nico90}
performed on the
orthopositronium (o-Ps) system --e$^+$e$^-$ bound in the $^3{\rm S}_1$
state-- have shown that its decay rate in vacuum is significantly
higher than the QED prediction. The data obtained by Westbrook
{\it et al.} \cite{Westbrook87&89}, give $\lambda_{exp}=7.0516\pm
0.0013\ \mu \second ^{-1} \label{exp}$, while Nico {\it et al.}
\cite{Nico90} obtain $\lambda_{exp}=7.0482\pm 0.0016\ \mu
\second ^{-1} \label{exp2}$. However,
the QED expectation \cite{Caswell79&Adkins83} is
$\lambda_{th}=7.03830\pm 0.00007\ \mu \second ^{-1}$,
so that there is a significant discrepancy between theory
and experiment.

One of the proposed solutions to this problem is that, apart from the
standard $3\gamma$ decay, there are new disintegration channels
of o-Ps. Indeed, if o-Ps has exotic decays, with a branching
ratio on the order of $10^{-3}$, the theoretical prediction would
increase in the right amount to be consistent with experiment.

Much attention has been devoted to decays with final states that
involve exotic
particles. The decay mode o-Ps $\rightarrow \gamma X$, where $X$ is a
weakly interacting particle \cite{Samuel88},
or the decay into ``invisible" final states,
could solve the orthopositronium decay rate puzzle.
Several groups have been searching for such new decays of o-Ps
\cite{Maeno95}--\cite{Akopyan91&Tsuchiaki90&Orito89&Atoyan89}.
The most restrictive experimental results on these exotic decays
that have been published are:

$\bullet$ o-Ps $\rightarrow \gamma X$ has, if $X$ is short-lived,
the following upper limits on the branching ratios:
$2.0 \times 10^{-4}$ for
$847\ \mbox{keV} \leq m_X \leq 1013\ \mbox{keV}$ \cite{Maeno95},
$3.0 \times 10^{-4}$ for
$m_X \leq 500\ \mbox{keV}$ \cite{Asai94}

$\bullet$ o-Ps $\rightarrow \gamma X$ has, if $X$ is long-lived,
a branching ratio that is less than $1.1 \times 10^{-6}$,
provided $m_X \leq 800\ \mbox{keV}$ \cite{Asai91}

$\bullet$ o-Ps $\rightarrow ``nothing"$, has also
been searched for.
The absence of any ``invisible" event gives a branching
ratio less than $2.8 \times 10^{-6}$ \cite{Mitsui93}.

We see that the decay o-Ps $\rightarrow ``nothing"$ is
ruled out as an explanation
of the o-Ps anomaly, and that the decay  o-Ps $\rightarrow \gamma X$
is excluded in most of
the possible $X$ mass range, $0 \leq m_X \leq 1022$ keV, but
{\sl not} in all the range.

The experimental limit on o-Ps $\rightarrow \gamma X$ is based on
the search for a monochromatic photon in the final state.
We note that a decay such as
o-Ps $\rightarrow \gamma X_1 X_2$, where there are now two
exotics in the final state
could provide a solution for the o-Ps puzzle, but
there are no experi\-mental
limits on such decays, the reason being that now the photon
is not monochromatic.

In this letter we place bounds on the o-Ps exotic decays
based on primordial nucleo\-synthesis arguments. We will obtain much
more restrictive bounds than the experimental results,
for {\sl any} decay of o-Ps into final
states containing at least one
exotic particle.

Simply stated, our idea is the following. A exotic particle
that is produced in o-Ps
decay is coupled to electrons, positrons and/or to photons with a
strength that will determine the
bran\-ching ratio for this mode. In the early Universe, that particle
will be maintained
in thermal equilibrium due to its exotic coupling to e$^+$, e$^-$
and/or $\gamma$.
However, the exotic
particle must decouple by the nucleosynthesis era ($T \sim 1$ MeV)
since otherwise the successful predictions of the standard primordial
nucleosynthesis would be spoiled. This requirement places strong
limits on the exotic coupling, and consequently on the exotic decay
of o-Ps.

We are interested in the contribution to the effective degrees
of freedom of new particles with masses up to the o-Ps mass,
{\it i.e.}, $0 \leq m \leq 1022$ keV. Therefore, in general,
we cannot neglect the mass $m$ compared to the
temperatures $T$ that we will have to
consider, since the latter go down to $T \sim 1$ MeV. We will discuss
how to treat this situation in Sect.2. In Sect.3 we will discuss
the $\gamma X$ decay channel, with $X$ a scalar particle, and in
Sect.4 we will discuss other decays. In Sect.5 we will give
our conclusions.\\

{\bf 2. The contribution of $X$ to the effective degrees of freedom}

As we will see, we will be mainly interested in the exotic particle $X$
being a scalar $S$. Let us then consider this case: $X=S$.
The scalar particle $S$ was in thermal equilibrium in the early
Universe, due to its
interaction with electrons, positrons and/or photons.
It has an interaction rate $\Gamma$ that can be
calculated once this interaction is specified.

As the Universe expands and cools, the interaction rate decreases
and $S$ decouples
at a time $t_D$ and at a temperature $T_D$ when
\begin{equation} \label{decou}
 \Gamma = H
\end{equation}
where $H$ is the expansion rate. Neglecting the curvature term,
a completely justified
approxi\-mation in the early Universe, $H$ is given by
\begin{equation} \label{H}
 H^2 =\frac{8\pi G}{3}\rho = g_*(T)\frac{8\pi^3 G}{90} \ T^4
\end{equation}
In (\ref{H}) we have defined the effective
degrees of freedom $g_*(T)$ (we follow the notation of Kolb
and Turner \cite{Kolb}).

As is well known, primordial nucleosynthesis offers a
limit on the degrees
of freedom $g_*$ contributing to the early Universe expansion at
$T\simeq 1$ MeV. It comes from a detailed comparison between the
observed abundances of the
primordial elements and the predictions of the Big Bang model
in the nucleosynthesis era \cite{Walker}.
The limit is very stringent and places a bound on the energy
density $\rho_S$ of $S$ compared to the photon
contribution $\rho_\gamma$\footnote{A very recent analysis \cite{Hata}
on primordial nucleosynthesis tends to suggest that the limit
is even smaller than (\ref{g*}). Our final upper bound on exotic
branching ratios would be then more stringent.}
\begin{equation} \label{g*}
 \Delta g_* \equiv \frac{\rho_S}{1/2\ \rho_\gamma}
            \leq 0.5  \ \ \ \ \ \ T \simeq 1\ \mbox{MeV}
\end{equation}

The scalar particle $S$ would of course contribute to $g_*$.
Its energy density is
\begin{equation} \label{rho}
\rho_S = \frac{1}{(2\pi)^3} \int dp\, 4\pi p^2 E f(p,t)
\end{equation}
where $E=\sqrt{p^2+m^2}$. Before decoupling, the particle
is in thermal equilibrium and its
distribution function $f$ is that of a boson at
temperature $T$
\begin{equation} \label{termal}
f(p,t) = [\exp(E/T)-1]^{-1} \ \ \ \ \ \ T \geq T_D
\end{equation}
The temperature $T$ is a decreasing function of $t$.
After decoupling, the number of $S$ particles
is conserved, which is expressed by
\begin{equation} \label{distr}
f(p,t) = f(p_D,t_D)   \ \ \ \ \ \ T < T_D
\end{equation}
where the momentum has been redshifted as $pR=p_D R_D$, with $R$
the scale factor.

{}From (\ref{termal}) and (\ref{distr}) we get the expression
\begin{equation} \label{nottermal}
f(p,t) =
        \left[\exp\sqrt{\frac{p^2}{\theta^2}+
        \frac{m^2}{T_D^2} }-1 \right] ^{-1}
        \ \ \ \ \ \ T < T_D
\end{equation}
where we have defined the time dependent parameter
\begin{equation} \label{def.theta}
\theta=\frac{R_D}{R}  T_D
\end{equation}
The parameter $\theta$ cannot be interpreted as the temperature
of $S$, unless $S$ is
massless. $\theta$ can be calculated using entropy conservation, which
involves the function
$g_{*S}(T)$, the effective degrees of freedom contributing
to the entropy,
\begin{equation} \label{entr.conserv.}
g_{*S}(T)R^3T^3 = g_{*S}(T_D)R_D^3T_D^3
\end{equation}
which gives
\begin{equation} \label{theta}
\theta= T \left( \frac{g_{*S}(T)}{g_{*S}(T_D)} \right)^{1/3}
\end{equation}
(Here $S$ refers to entropy).

We have evaluated $\Delta g_*$, as defined in (\ref{g*}), numerically,
substituting (\ref{nottermal}) and (\ref{theta}) into (\ref{rho})
and using the values of $g_*(T)$
and  $g_{*S}(T)$ calculated using their exact definitions
(see for example \cite{Kolb}). Then, in order not to spoil
the agreement of the predictions of the Hot Big Bang model
with the observed primordial abundances, we impose the experimental
limit (\ref{g*}). In the mass range $0\leq m \leq 1022$ keV, the limit
translates into the condition that the $S$ species must decouple
for all the temperatures in the range
\begin{equation} \label{TD}
1 \ \mbox{MeV} \leq T \leq 100 \ \mbox{MeV}
\end{equation}

{\bf 3. The o-Ps $\rightarrow \gamma X$ decay, with $X=S$ a
scalar particle}

We will assume in this section that $C$ and $P$ are conserved.
There are three ways in which this decay can then proceed:

(A) $S$ is coupled to electrons. (An equivalent
model with $X$ a pseudoscalar particle was considered by Samuel
\cite{Samuel88}.)
The o-Ps decays via
Fig.\,(1.A).  The exotic coupling is described by a
Yukawa-type interaction
\begin{equation} \label{lagr1}
 {\cal L}_A = g_A\, \bar \psi \psi\, S
\end{equation}
where both here and in the following $\psi$ is the electron field.

(B) $S$ is coupled to photons. The decay proceeds through the diagram
Fig.\,(1.B). The effective Lagrangian describing the new coupling is
\begin{equation} \label{lagr2}
 {\cal L}_B = \frac{1}{4}\, g_B\, F^{\mu \nu} F_{\mu \nu}\, S
\end{equation}

(C) There is a contact term coupling electrons to a photon and
$S$ as in Fig.\,(1.C). The Lagrangian describing the interactions has
the gauge-invariant form
\begin{equation} \label{lagr3}
 {\cal L}_C = \frac{1}{2}\, g_C\, \bar \psi
              \sigma^{\mu \nu }\psi\, F_{\mu \nu}\, S
\end{equation}

Let us start with model (A). We have an exotic particle $S$ of
mass $m_S$, with the interaction  Lagrangian given in (\ref{lagr1}).

In the early Universe, the scalar $S$ is in thermal equilibrium
due to the processes
\begin{eqnarray}
\gamma\, S \ & \rightarrow & \  e^+e^-  \label{photonS} \\
 e\, S \ & \rightarrow & \ e\, \gamma   \label{eS}
\end{eqnarray}
where $e$ can be either $e^+$ or $e^-$.
We have to consider all three processes since their contribution
to the interaction rate are of approximately the same magnitude.

Let us work out the contribution of the process
$\gamma S \rightarrow e^+e^-$ in some detail. We are interested
in temperatures $3\, T \geq 3\, T_D >> m_e,m_S$. In this limit,
the process has a total cross section given by
\begin{equation} \label{sigma}
 \sigma[s] =\frac{\alpha g_A^2}{s} \ln s/m_e^2
\end{equation}

The corresponding interaction rate (per unit time and per $S$) is
\begin{equation} \label{int.rate}
 \Gamma = n_\gamma  <\sigma v>
\end{equation}
 where the number density of photons is given by
\begin{equation} \label{dens.ph}
 n_\gamma=  2 \frac{\xi(3)}{\pi^2} T^3
\end{equation}
The thermalized quantity $<\sigma v>$ appears in the evolution
equation of the $S$ number density $n_S$ (the superscript $eq$
refers to the distribution in thermal equilibrium)
\begin{equation} \label{Boltzmann}
 \dot{n}_S  + 3Hn_S = - <\sigma v> n^{eq}_\gamma [n_S-n^{eq}_S]
\end{equation}
One can find approximate expressions for $<\sigma v>$ in
\cite{Bernstein/Srednicki/Gelmini}.
We write the expression for this quantity in the relativistic
limit we are interested in
\begin{equation} \label{integral}
 <\sigma v> =\frac {1}{n^{eq}_\gamma n^{eq}_S}
             \int\ dn^{eq}_\gamma\ dn^{eq}_S\
             (1-\cos\theta) \ \sigma[2E_\gamma E_S
             (1-\cos\theta)]
\end{equation}
Here $E_i$ refers to the energy of the particle $i$ and $\theta$ is
the angle between $S$ and $\gamma$. Substituting now the cross
section (\ref{sigma}) into this expression,
one obtains after a straightforward calculation
\begin{equation} \label{th.av.}
<\sigma v> =\frac{\pi^4}{288 \ \xi (3)^2} \frac{\alpha g_A^2}{T^2}
             \left\{ \ln 4 -\frac{1}{4}+2\left( 1-\gamma +\frac{6
             \dot{\xi} (2)}{\pi^2}
             \right) + 2 \ln \frac{T}{m_e} \right\}
\end{equation}
where $\xi (x)$ is the Riemann $\xi$ function, $\dot{\xi} (x)$ its
derivative and $\gamma$ is the Euler constant.

We can now obtain restrictions on $g_A$ by requiring that $S$
is decoupled in the temperature range determined in (\ref{TD}).
For each $T_D$ in this range, we
obtain an upper limit on $g_A$, and our final result
is necessarily the
{\sl most} stringent of them all. It turns out that this numerical
limit is obtained at $T_D^A\simeq 1$ MeV.

The formulae above actually only take account of the process
(\ref{photonS}), which is one of the
reactions that maintains $S$ in equilibrium. The contributions of
the other processes (\ref{eS})
can be worked out along the same lines. If one takes
all these processes into account, the upper bound on the coupling $g_A$
appearing in the effective Lagrangian (\ref{lagr1}) is found to be
\newpage
\begin{equation} \label{bound}
g_A \leq 5 \times 10^{-10} \equiv g_A^{upper}
\end{equation}
where we have defined (for a later purpose) the numerical value
of the upper limit as $g_A^{upper}$.

With the Lagrangian (\ref{lagr1}), one can calculate
the branching ratio
\begin{equation} \label{B_S}
 B_S = \frac{\Gamma(\mbox{oP-s}\rightarrow \gamma S)}
       {\Gamma(\mbox{oP-s})} =
       g_A^2 \ \frac{3}{8(\pi^2 -9) \alpha^2} \
       \left( 1-\frac{m_S^2}{4m_e^2} \right)
\end{equation}

This leads directely to one of our main results. Inserting (\ref{bound})
in (\ref{B_S}), the following stringent limit is found
\begin{equation} \label{limit}
B_S \leq 2 \times 10^{-15} f_A(m_S) \leq 2 \times 10^{-15} \equiv  B_{upper}^A
\end{equation}
where the function $f_A(m_S) = B_S(m_S)/B_S(m_S=0)= 1-
m_S^2/4m_e^2 \leq 1$, with  $B_S$ calculated
using (\ref{lagr1}).

As we discussed at the beginning of this section, the decay
o-Ps $\rightarrow \gamma S$ is due to new interactions that can
be described by effective Lagrangians. We have presented the
consequences of assuming that the exotic decay is due
to the new interaction (A).
Using (\ref{lagr1}) we obtained the limit (\ref{limit}) on $B_S$. We
have performed similar analyses in the cases (B) and (C), using
(\ref{lagr2}) and (\ref{lagr3}) respectively. We will only quote the
value of the limits.

For the interactions (B), we obtain
\begin{equation} \label{limit.gB}
g_B \leq 2 \times 10^{-7} \ \mbox{GeV}^{-1}  \equiv g_B^{upper}
\end{equation}
and
\begin{equation} \label{limit.BB}
 B_S \leq 2 \times 10^{-17} f_B(m_S) \leq 2 \times 10^{-17} \equiv B_{upper}^B
\end{equation}
with $f_B(m_S)= \left( 1-m_S^2/4m_e^2 \right) ^3 \leq 1$. The
limit is now obtained at $T_D^B \simeq 100$ MeV.

For the case (C), we obtain
\begin{equation} \label{limit.gC}
g_C \leq 1 \times 10^{-7} \ \mbox{GeV}^{-2}   \equiv g_C^{upper}
\end{equation}
and
\begin{equation} \label{limit.BC}
 B_S \leq 1 \times 10^{-22} f_C(m_S) \equiv  B_{upper}^C
\end{equation}
with $f_C(m_S)= \left( 1-m_S^2/4m_e^2 \right) ^3 \leq 1$.
The limit is also obtained at $T_D^C \simeq 100$ MeV.

The bound on $B_S$ we finally can claim is the least
restrictive, ${\sl i.e.}$, $B_S \leq 2 \times 10^{-15}$.\\

{\bf 4. Other o-Ps decays to exotic particles}

In order to discuss other decays, it will be very useful
to understand why we have found that the limit on $B_S$
is less restrictive in case (A) than in (B), and why the
latter is less restrictive than case (C).

The differences arise from the different dimensions $D$ of the
operators appearing in the
effective Lagrangians. The operator in (\ref{lagr1}) has $D=4$,
in (\ref{lagr2}) it has $D=5$, while
in (\ref{lagr3}) it has $D=6$. Consequently, the coupling $g_A$ is
dimensionless, while $g_B$ and
$g_C$ have dimensions $M^{-1}$ and $M^{-2}$, respectively.
It is important to notice that there are two energy scales
in our problem. On the one hand, we obtain
restrictions at temperatures $T_D$, where decoupling is necessary in
order not to excessively modify the predicted abundances in the
primordial nucleosynthesis. On the other hand, we use
these restrictions
into the exotic o-Ps decay, that occurs at an energy scale
$E \sim m_e$. This leads to the
following relations among the upper limits on the exotic couplings
and on the branching
ratios, that are valid up to an order of magnitude:
\begin{equation} \label{relations}
B_{upper}^A \sim \frac{T_D^A T_D^B}{m_e^2} B_{upper}^B\ \ , \ \ \ \
B_{upper}^B \sim \frac{T_D^B T_D^C}{m_e^2} B_{upper}^C
\end{equation}

We can understand from this discussion that the least restrictive
limit on $B_S$ is in the
case that the coupling is dimensionless, as it is in (A).
This discussion will be the
key to show the main point of this section, namely that all
decays involving
exotic particles in the final state have {\sl at least}
the upper limit $B_{upper}^A$, corresponding to case (A) above.

The o-Ps $\rightarrow \gamma X$ decay, with $X$ a pseudoscalar
particle $P$, is for our
purposes completely similar to the decay where $X$ is
a scalar particle. There are also three ways in which the decay
can proceed, described by the three Lagrangians
\begin{eqnarray}
{\cal L}_1 & = & g_1 \, \bar \psi \gamma_5 \psi\, P  \\
{\cal L}_2 & = & \frac{1}{8}\, g_2 \,
               \epsilon_{\alpha \beta \gamma \delta}
               F^{\alpha \beta} F^{\gamma \delta}\, P  \\
{\cal L}_3 & = & \frac{1}{2}\, g_3 \, \bar \psi  \sigma^{\mu \nu }
               \gamma_5 \psi\, F_{\mu \nu}\, P
\end{eqnarray}
The conclusions regarding the upper limits one can obtain on $B_P$
are {\sl exactly} the same as the corresponding cases (A), (B)
and (C) above. Thus $B_{upper}^A$ is the least restrictive upper limit
we can place on the branching ratio of the decay
o-Ps $\rightarrow \gamma P$. One can think
of interactions that do not conserve C and/or P and leading to o-Ps
$\rightarrow \gamma X$, with $X$ a scalar or pseudoscalar particle.
The results one finds are again {\sl exactly} the upper limits that
we found in cases (A), (B) and (C).

Let us now consider the decay o-Ps $\rightarrow \gamma X$,
with $X$ a spin one or two particle. The effective
Lagrangian describing the exotic interaction contains operators
with $D \geq 4$. The least restrictive limit will be obtained
when $D=4$, since then the coupling is dimensionless. In fact,
since $X$ now has more degrees of freedom that could contribute
at the nucleosynthesis era, the decoupling temperature that we
obtain is higher than the one we found in the case of only one
degree of freedom  ($X=S$ or $P$). Therefore the limit we
obtain is even
several times more restrictive than that of case (A).
Thus $B_{upper}^A$ has to be seen as a conservative upper limit
when $X$ has spin different from zero.

Another decay that has been experimentally investigated is o-Ps
$\rightarrow ``nothing"$. A possibility here is o-Ps
$\rightarrow S V$,
with both the scalar $S$ and the vector $V$ independently attached
to electrons. Here we have two effective Lagrangians with $D=4$,
and two independent dimensionless couplings:
\begin{eqnarray} \label{attachedtoe}
{\cal L}_4 & = & g_4\, \bar \psi \psi\, S \\
{\cal L}_5 & = & g_5\, \bar \psi \gamma^\mu \psi\, V_\mu
\end{eqnarray}
Our arguments
apply to the product of these two couplings. We have to take into
account that one has now two exotic particles and thus more degrees
of freedom, and the decoupling will have to occur at higher
temperatures making the bound on the branching ratio
$\Gamma(\mbox{o-Ps}\rightarrow
{\rm ``nothing"})/ \Gamma(\mbox{o-Ps})$
even more stringent than $B_{upper}^A$.
Another possibility is to have contact terms like
\begin{eqnarray} \label{contact}
{\cal L}_6 & = & g_6\, \bar \psi \gamma^\mu \psi\, V_\mu\, S \\
{\cal L}_7 & = & g_7\, \bar \psi \sigma^{\mu \nu} \psi
                 \, (\partial_\mu V_\nu ) \, S
\end{eqnarray}
The coupling $g_6$ has dimension $M^{-1}$ and the coupling $g_7$
has dimension $M^{-2}$; thus the bound on the branching
ratio will be much more
restrictive than $B_{upper}^A$. There are still other ways to obtain a
o-Ps $\rightarrow ``nothing"$ decay. We will not list all the
possibilities, since it is easy to see that we will always
have limits
more stringent than $B_{upper}^A$.

An interesting decay of o-Ps would be into a photon and {\sl two} exotic
particles. In fact, experiments are optimized to search for a
monochromatic photon, so that they are not sensitive to this type
of decay. We can apply our arguments based on nucleosynthesis
to restrict this possibility. For example, consider o-Ps
$\rightarrow \gamma S_1 S_2$. The case that the two scalars are
independently attached to electrons is similar to the above mentioned
o-Ps $\rightarrow S V$ decay, and one reaches the same conclusion,
namely that $B_{upper}^A$ is a conservative upper limit on the
branching ratio of the exotic decay. If both scalar
particles emerge in a contact term with electrons, or if the scalars
are coupled to two photons, the coupling is dimensionful and the
bound is still more stringent.

The type of arguments discussed in this section can be applied
to {\sl any} decay with exotic particles in the final state.
One first has to write the effective Lagrangians that can lead
to the decay channel. The order of magnitude of the limit on
the branching ratio depends on the dimensions of the coupling,
and to a lesser extent on the number of exotic degrees of
freedom. The least stringent limit is obtained when the
coupling is dimensionless and there is a scalar or pseudoscalar
particle in the final state (model (A) or closely related),
so that we place the following limit on {\sl any} exotic
decay
\begin{equation}
B = \frac{\Gamma(\mbox{o-Ps}\rightarrow
    {\rm``exotic"} + ...)}{\Gamma(\mbox{o-Ps})}
    \leq 2 \times 10^{-15}
\end{equation}
\\

{\bf 5. Conclusions}

The o-Ps decay puzzle would be solved if there were new decays
with branching ratios $\sim 10^{-3}$. Some of these decays
have been ruled out by existing laboratory experiments.

In this letter we have shown that all o-Ps decays involving
exotic particles are restricted by
nucleosynthesis arguments.
Using an effective Lagrangian approach, we have been able
to place a very stringent upper limit for {\sl any} exotic decay
\begin{equation} \label{exotic}
B = \frac{\Gamma(\mbox{o-Ps}\rightarrow
    {\rm``exotic"} + ...)}{\Gamma(\mbox{o-Ps})}
    \leq 2 \times 10^{-15}
\end{equation}\\
Our result excludes the exotic decay solution to the
o-Ps puzzle.

{\bf Acknowledgements}

We thank the Theoretical Astroparticle Network for support under
the EEC Contract No. CHRX-CT93-0120 (Direction Generale 12 COMA).
This work has been partially supported by the CICYT Research Project
No's. AEN-93-0474 and AEN-93-0520.
R.E. acknowledges a FPI Grant from the Universitat Aut\`onoma de
Barcelona and
R.T. acknowledges a FPI Grant from the Ministeri d'Educaci\'{o}
i Ci\`{e}ncia (Spain).

\newpage

{\bf Figure Caption}

{\bf Fig.1.} Exotic o-Ps decays due to the effective Lagrangians
in models (A), (B) or (C).

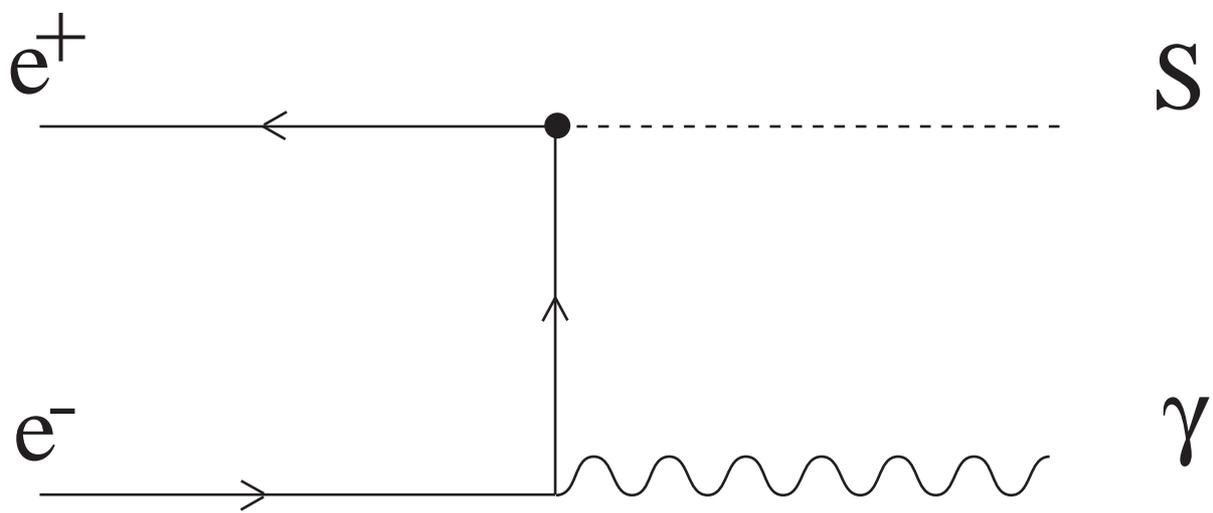
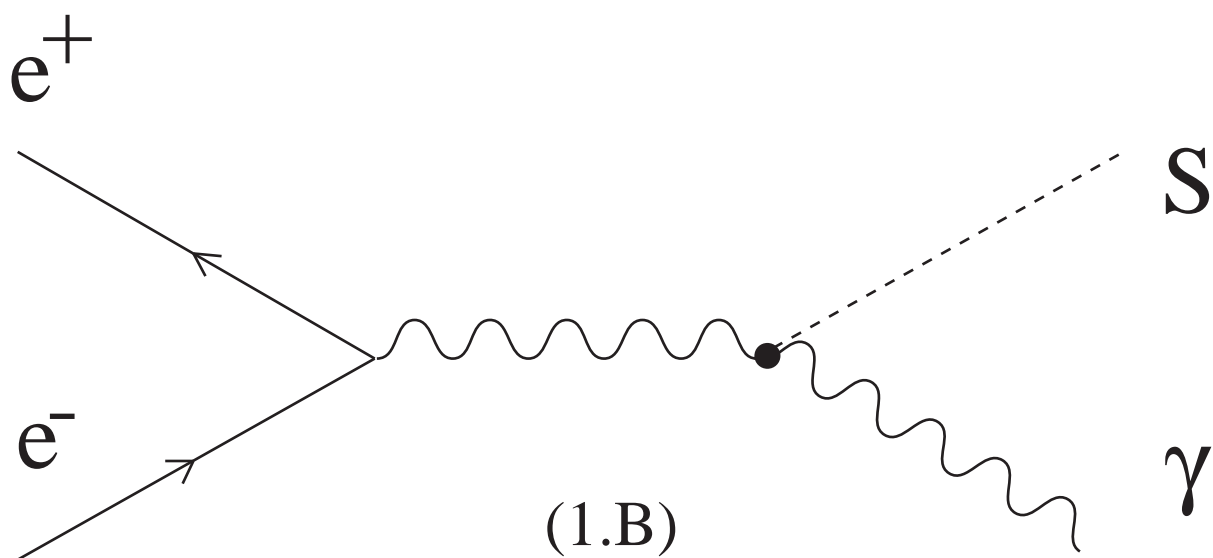
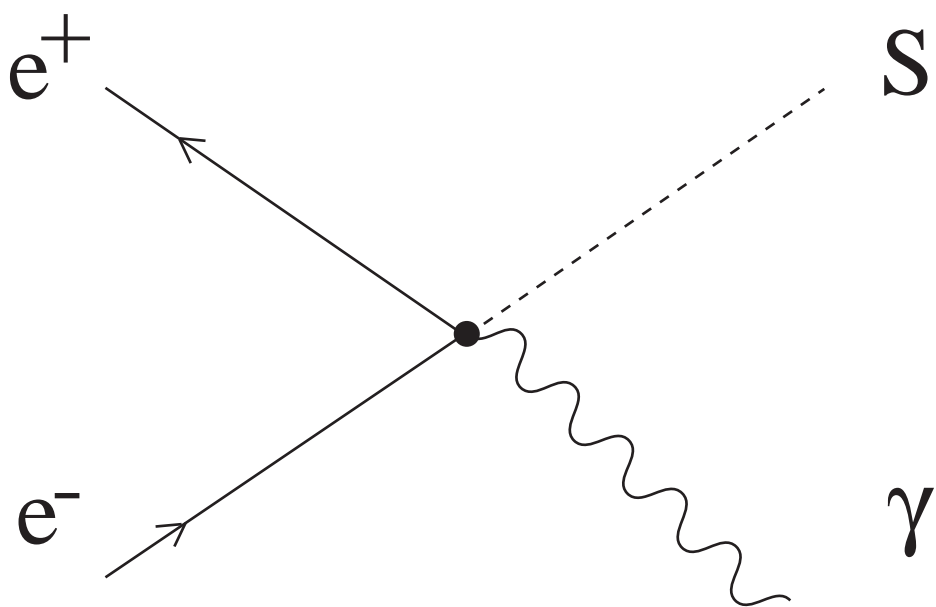
\end{document}